\documentclass[a4paper,twoside]{article}

\usepackage{epsfig}
\usepackage{subcaption}
\usepackage{calc}
\usepackage{amssymb}
\usepackage{amstext}
\usepackage{amsmath}
\usepackage{amsthm}
\usepackage{multicol}
\usepackage{pslatex}
\usepackage{apalike}
\usepackage{algorithm2e}
\usepackage[bottom]{footmisc}
\usepackage{SCITEPRESS}
\usepackage{tikz}
\usetikzlibrary{quantikz2}
\usepackage[hidelinks]{hyperref}

\usepackage{adjustbox}
\usepackage{tabularx}
\usepackage[capitalise]{cleveref}
\creflabelformat{equation}{#2#1#3}
\newcolumntype{C}{>{\centering\arraybackslash}X}

\begin{document}

\title{A Reinforcement Learning Environment for Directed Quantum Circuit Synthesis}

\author{\authorname{Michael Kölle\sup{1}, Tom Schubert\sup{1}, Philipp Altmann\sup{1}, Maximilian Zorn\sup{1}, Jonas Stein\sup{1} and Claudia Linnhoff-Popien\sup{1}}
\affiliation{\sup{1}Institute of Informatics, LMU Munich, Munich, Germany}
\email{michael.koelle@ifi.lmu.de}
}

\keywords{Reinforcement Learning, Quantum Computing, Quantum Circuit Synthesis}

\abstract{
With recent advancements in quantum computing technology, optimizing quantum circuits and ensuring reliable quantum state preparation have become increasingly vital. Traditional methods often demand extensive expertise and manual calculations, posing challenges as quantum circuits grow in qubit- and gate-count. Therefore, harnessing machine learning techniques to handle the growing variety of gate-to-qubit combinations is a promising approach. In this work, we introduce a comprehensive reinforcement learning environment for quantum circuit synthesis, where circuits are constructed utilizing gates from the the Clifford+T gate set to prepare specific target states. 
Our experiments focus on exploring the relationship between the depth of synthesized quantum circuits and the circuit depths used for target initialization, as well as qubit count. We organize the environment configurations into multiple evaluation levels and include a range of well-known quantum states for benchmarking purposes. We also lay baselines for evaluating the environment using Proximal Policy Optimization. By applying the trained agents to benchmark tests, we demonstrated their ability to reliably design minimal quantum circuits for a selection of 2-qubit Bell states.
}

\onecolumn \maketitle \normalsize \setcounter{footnote}{0} \vfill

\section{INTRODUCTION} \label{sec:introduction}
The field of quantum computing, including quantum sensing, quantum meteorology, quantum communication, and quantum cryptography is recently receiving a lot of attention~\cite{Toth2014,Ekert1991}. Consequently, directed quantum circuit synthesis (DQCS) involving quantum state preparation, which plays a vital role in the above-mentioned technologies, gains more and more interest. Current quantum circuit layouts range from rather straightforward ones, as used for Bell state preparation~\cite{Barenco1995,Bennett,Nielsen2010}, to highly sophisticated designs including tunable parameters, as present in Variational Quantum Classifiers and Variational Quantum Eigensolvers~\cite{Farhi2018,Schuld2018b,Peruzzo2013}. Though many approaches addressing quantum circuit development are known, most of them focus on the optimization of already existing circuits~\cite{Foesel2021,Zikun2023}. On the contrary, research regarding the circuit synthesis is sparse, making manual methods still a state-of-the-art technique for tackling these tasks. This circumstance gets especially problematic if the involved quantum circuits increase in qubit number and gate count, resulting in large state-spaces with an exponential amount of possible layouts. Hence, to guarantee efficient task-solving, it is crucial to develop an approach that tackles the problem of DQCS and decreases the amount of required human insight.

On consideration of the aforementioned examples, it becomes apparent that machine learning (ML) approaches are especially suited for subjects involving complex calculations and elaborate combinatorial optimizations. Hence, we consider a ML-based technique for the DQCS problem. Since quantum circuit construction and optimization does not entail learning from data in the classical ML sense, we instead apply the concept of improving by repeated interaction with a problem environment via reinforcement learning (RL). While research for applying RL on state preparation tasks is known, there is limited exploration of the underlying DQCS problem and the implementation on actual quantum hardware utilizing a set of distinct quantum gates (cf.~\cite{Gabor2022,Mackeprang2019}). Similarly, the disassembly of a proposed circuit into a sequence of valid quantum gates is a crucial step in facilitating the transfer to a real quantum device~\cite{mansky2022nearoptimal}.

We introduce the Quantum Circuit RL environment designed to train RL agents on the task of preparing randomly generated quantum states utilizing the Clifford+T gate set, enabling the trained agents solve the DQCS problem for arbitrary target states. In our environment, we view the use of quantum gates on quantum states as actions and the quantum state as observation. The objective is to prepare a specific target state efficiently, and success is measured by minimizing the number of gates needed to construct the quantum circuit. We also evaluate Proximal Policy Optimization algorithm agents on different configurations of our environment to form a baseline. Lastly, we benchmark the trained agents on a set of well-known 2-qubit states.
\section{RELATED WORK} \label{sec:related-work}
In this section we introduce underlying concepts essential for a clear comprehension of our research. Further we provide an overview of prior work in the field of AI-assisted quantum circuit generation and optimization, connecting it to our approach.

\subsection{Quantum States}
Quantum computing is an emerging technology distinguishing itself from classical computing fundamentally by the usage of so-called quantum bits or qubits instead of classical bits as units of information storage. Qubits among other aspects differ from their classical counterparts by showing the ability of superposition. This describes the ability of the qubit to not only be in one of the discrete states 0 or 1, like a classical bit, but to be in any linear combination of 0 and 1, enlarging the available state-space from discrete to continuous. \cref{eq: qubitstate} defines the quantum state of one qubit in Dirac- and vector-notation.

\begin{align}
    \ket{q} = \alpha \ket{0} + \beta \ket{1} = \begin{pmatrix}
        \alpha \\
        \beta
    \end{pmatrix}\; \; \; \; \; \;\alpha, \beta \; \in \;\mathbb{C}\label{eq: qubitstate}
\end{align}
Another characteristic of qubits is the so-called entanglement, which describes the possibility of correlating the states of multiple qubits, enabling the setup of complex relations between them.
Another option for the description of quantum states is the density matrix representation given in \cref{eq: qubitmatrix}. 

\begin{align}
    \rho = \ket{q}\bra{q} = \begin{pmatrix}
        |\alpha|^2 & \alpha\beta^{*}\\
        \alpha^{*}\beta & |\beta|^2
    \end{pmatrix}\; \; \; \; \; \;\alpha, \beta \; \in \;\mathbb{C}\label{eq: qubitmatrix}
\end{align}
To compare two density matrices and hence two quantum states $\rho$ and $\widetilde{\rho}$, the fidelity $F$ as given in \cref{eq: fidelity} is used as a measure.

\begin{align}
    F(\rho,\widetilde{\rho}) = \biggl(tr\biggl(\sqrt{\sqrt{\rho}\widetilde{\rho}\sqrt{\rho}}\biggr)\biggr)^{2} \label{eq: fidelity}
\end{align}
When solely states in vector representation are considered, this equation simplifies to the expression given in \cref{eq: purefidelity}\cite{Jozsa1994}.

\begin{align}
    F(\rho,\widetilde{\rho}) = |\bra{q}\ket{\widetilde{q}}|^2 \label{eq: purefidelity}
\end{align}
\subsection{Quantum Circuits}
Quantum computers use logic gates, represented by unitary matrices denoted as $U$, to transform a quantum state $\ket{\Phi}$ into a new state $\ket{\Phi'}$ following \cref{eq: quantumtrafo}.

\begin{align}
    \ket{\Phi'} = U \ket{\Phi} \label{eq: quantumtrafo}
\end{align}
The unitarity aspect of the gates makes all operations relying on these gates unitary and hence reversible. A generic unitary matrix working as a quantum mechanical operator on a single qubit can be defined according to \cref{eq: genericunitary}.

\begin{align}
    U = e^{i\frac{\gamma}{2}}\begin{pmatrix}
        \cos{\theta} e^{i\rho}\; \;\;\;\;\;\;\;\;\sin{\theta} e^{i\phi} \;\\
        \;-\sin{\theta} e^{-i\phi} \;\;\;\; \cos{\theta} e^{-i\rho}\;
        \end{pmatrix} \label{eq: genericunitary} 
\end{align}  
In the unitary matrix \( U \), \( \gamma \) denotes a global phase multiplier, \( \theta \) characterizes the rotation between computational basis states, while \( \rho \) and \( \phi \) introduce relative phase shifts to the diagonal and off-diagonal elements respectively. Different quantum gates can be created by selecting specific values for the involved parameters. By leaving some of the involved parameters undefined, the design of parameterized quantum gates is possible. Another characteristic of a quantum gate is its multiplicity, defining the number of qubits the gate acts on. Several different gates can then be combined within a gate set (e.g. the strictly universal Clifford+T gate set given in \cref{table : gatesets}).\newline  
\begin{table}[htbp]
\begin{center}
\adjustbox{max width=\linewidth}{
    \begin{tabular}{ |c|c|c| } 
        \hline
         Symbol (Name) & Matrix representation & Circuit notation\\ 
        \hline\hline
        \rule{0pt}{20pt} I (Identity)  &  $ \begin{pmatrix} 1 & 0 \\ 0 & 1 \\ \end{pmatrix} $ & \begin{quantikz}[thin lines]&\gate{I}&\end{quantikz}\\[10pt]
        \hline
        \rule{0pt}{20pt}H (Hadamard)  &  $ \frac{1}{\sqrt{2}}\begin{pmatrix} 1 & 1 \\ 1 & -1 \\ \end{pmatrix} $ & \begin{quantikz}[thin lines]&\gate{H}&\end{quantikz}\\[10pt]
        \hline
        \rule{0pt}{20pt}S   &  $ \begin{pmatrix} 1 & 0 \\ 0 & e^{i \frac{\pi}{2}} \\ \end{pmatrix} $ & \begin{quantikz}[thin lines]&\gate{S}&\end{quantikz}\\[10pt]
         \hline
        \rule{0pt}{30pt}CNOT   &  $ \begin{pmatrix} 1 & 0 & 0 & 0\\ 0 & 1 & 0 & 0\\0 & 0 & 0 & 1\\0 & 0 & 1 & 0\\ \end{pmatrix} $& \begin{quantikz}[thin lines]&\ctrl{1}&\\&\targ{}&\end{quantikz}\\[15pt]
         \hline
        \rule{0pt}{20pt}T   &  $ \begin{pmatrix} 1 & 0 \\ 0 & e^{i \frac{\pi}{4}} \\ \end{pmatrix} $ & \begin{quantikz}[thin lines]&\gate{T}&\end{quantikz}\\[10pt]
         \hline
    \end{tabular}
    }
    \caption{The gates in the Clifford+T gate set, along with their corresponding matrix representations and circuit symbols. For the CNOT gate, the example provided illustrates the scenario where the first qubit acts as the control bit.}
    \label{table : gatesets}
\end{center}
\end{table}
A quantum circuit is then formed by a sequence of quantum gates acting upon a number of qubits, while the overall number of gates within the quantum circuit is called the circuit-depth. Hence the circuit represents one big transformation matrix, transforming the incoming quantum state provided by the input qubits into an altered quantum state obtained on the output qubits.

\subsection{Quantum State Preparation} \label{subsec: quantumstateprep}
Recently much effort has gone into the investigation of quantum state preparation for different quantum-based fields like quantum meteorology, quantum sensing, quantum communication, and quantum computing to name just a few examples~\cite{Krenn2015,Krenn2020a,Mackeprang2019}.
Further exploiting computational resources, approaches facilitating state preparation using ML methods were done Mackeprang et al. investigated the state preparation of 2-qubit quantum states using RL algorithms showing their ability to generate Bell states finding the same solutions as previously discovered by humans~\cite{Mackeprang2019,Zhang}.

A different approach towards automated quantum state preparation was investigated by Gabor et al.~\cite{Gabor2022}. Their study involved an approach training an RL agent to generate quadratic transformation matrices transforming a initial state into a target state, both given in the density matrix representation. 
To account for unitarity of the transformation, they utilized a QR-decomposition disassembling the initially obtained matrix A according to $A = U\cdot R$, while $U$ represents a unitary and $R$ an upper triangular matrix. Subsequently, they used $U$ as the unitary transformation matrix. With this technique they reached state fidelities of up to $> 0.99$ for individual target-states, but struggled with arbitrary state preparation, resulting in unrealistically long circuits (even on small 2 qubit circuits)~\cite{Gabor2022}. 

Their research showed a promising way to approach the subject of quantum state preparation opening up new possibilities, but also pointed out several difficulties. The studies done in this project are closely connected and partially based on the research of Gabor et al., dealing with a similar kind of state preparation problem, while focusing on the possible improvements in the following. One issue of the approach chosen by Gabor et al. might be the construction of potentially non-unitary quadratic matrices, leading to costly QR-decomposition scaling with a complexity between $\mathcal{O}(n^{2})$ and $\mathcal{O}(n^{3})$, if $n$ is the dimension of the quadratic matrix~\cite{Parlett2000}.
To improve this approach, we exclusively utilize unitary transformations as provided by the Clifford+T gate set, reducing computational demands and accelerating agent learning. 
To simplify the approach used by Gabor et al. we substituted the density matrix representation by a vector representation, reducing the dimensions of the representation from $N^2$ to $N$ potentially lowering the computational costs. 

\subsection{Quantum Circuit Optimization and Synthesis}
Since manual optimization is a time-consuming, error-prone process requiring a high amount of knowledge, automation of quantum circuit synthesis and optimization is crucial~\cite{Foesel2021}. Consequently, there is a rising focus on employing machine learning algorithms to tackle this challenge~\cite{Cerezo2020,Pirhooshyaran2020,Ostaszewski2021,altmann2023challenges}.
An approach from Fösel et al. aims to optimize arbitrary generated quantum circuits  with regards to their complexity using a CNN approach, yielding an overall depth and gate reduction of 27\% and 15\% respectively~\cite{Foesel2021}. 
Further Zikun et al. implemented an RL-based procedure utilizing a graph-based framework to represent the structure of a certain quantum circuit. Their proposed algorithm (QUARL) then optimizes the respective circuit with regard to its gate count, while maintaining its overall functionality achieving a gate reduction ranging around 30\%. However, most of the approaches focus on the optimization of already existing circuits disregarding their initial synthesis. To address this issue, our study includes the initial quantum circuit synthesis into the procedure~\cite{Zikun2023,Xu2022}.

\clearpage
\section{QUANTUM CIRCUIT ENVIRONMENT} \label{sec:approach}

In this section, we introduce a versatile and scalable reinforcement learning environment designed for quantum circuit synthesis. This environment establishes a foundational platform for researchers to employ machine learning in discovering new and efficient quantum circuits for known problems. At each step, a RL agent can place one quantum gate onto a circuit, with the aim of crafting a circuit that maps from an arbitrary initial state to an arbitrary target state. 
We formulate the problem at hand as a \textit{Markov decision process} $M = \langle S, \Gamma, \mathcal{P}, R, \gamma \rangle$  where $S$ is a set of states $s_t$ at time step $t$, $\Gamma$ is a set of actions $a_t$ , $\mathcal{P}(s_{t+1}|s_{t},a_{t})$ is the transition probability from $s_t$ to $s_{t+1}$ when executing $a_t$, $r_t = R(s_t,a_t)$ is a scalar reward, and $\gamma \in [0,1)$ is the discount factor~\cite{puterman2014markov}.
For our implementation, we use the Pennylane framework to efficiently simulate quantum circuits. The project is open-source\footnote{\href{https://github.com/michaelkoelle/rl-qc-syn}{https://github.com/michaelkoelle/rl-qc-syn}}, distributed under the MIT license and available as a package on PyPI. In the following sections we elaborate on the details of the Quantum-Circuit environment. 


\subsection{Observation Space}
We define an observation in our environment as a real vector of length $2^{n+2}$. Since a normalized \(n\)-qubit quantum system can be expressed as a complex vector in \(2^n\) dimensions, we start with two complex vectors describing the current quantum state $(v)$ and the desired target quantum state $(\hat{v})$, each of length $2^n$. 

\begin{equation}
\small
\begin{split}
    v &= \begin{pmatrix}
        v_{1}\\
        v_{2}\\
        \cdot\\
        \cdot\\
        \cdot\\
        v_{2^n}
    \end{pmatrix} \; \cap \; \hat{v} = \begin{pmatrix}
        \hat{v}_{1}\\
        \hat{v}_{2}\\
        \cdot\\
        \cdot\\
        \cdot\\
        \hat{v}_{2^n}
    \end{pmatrix} \; \Rightarrow{} \; s = \begin{pmatrix}
        \operatorname{Re}(v_{1})\\
        \operatorname{Re}(v_{2})\\
        \vdots\\
        \operatorname{Re}(v_{2^n})\\
        \operatorname{Im}(v_{1})\\
        \operatorname{Im}(v_{2})\\
        \vdots\\
        \operatorname{Im}(v_{2^n})\\
        \operatorname{Re}(\hat{v}_{1})\\
        \operatorname{Re}(\hat{v}_{2})\\
        \vdots\\
        \operatorname{Re}(\hat{v}_{2^n})\\
        \operatorname{Im}(\hat{v}_{1})\\
        \operatorname{Im}(\hat{v}_{2})\\
        \vdots\\
        \operatorname{Im}(\hat{v}_{2^n})
    \end{pmatrix} \\
    &\forall_{i \in\{1 \cdot\cdot 2^n\}} \; v_{i},\hat{v}_{i} \in \mathbb{C} \;\;\cap   \;\; s  \in  S    
\end{split}
\label{eq: obsspace}
\end{equation}

The concatenation of both results in a complex vector of length $2^{n+1}$. Splitting up the complex coefficients of the resulting vector into the real and imaginary part we end up with the $2^{n+2}$ real dimensions characterizing the vector \(s\) in observation-space $S$. With the current state $(v)$ included in \(s\) describing the complete state of the quantum system, the environment is fully observable. Further the observation contains the desired target state $(\hat{v})$ to maintain a consistent target perspective for the RL agent.

\subsection{Action Space}

The action space \(\Gamma\) is multi-discrete and defined by two finite sets of a specific size, out of which one element is drawn respectively to form an action. While one set accounts for all gates included in the input gate set \(G\), the other set represents all possible combinations of qubits inputted into the respective gate, given a total number of \(n\) qubits. \cref{eq: actspace} calculates the size of the action-space with regards to the given \(G\) and \(n\), with \(n_{\text{max}}\) being the number of qubits taken by the gate \(g\in G\) processing the highest number of qubits within the gate set.

\begin{align}
\begin{split}
    \Gamma = \left[\{0,1..,|G|\},\{0,1..,C\}\right]\\
    \text{with} \quad  C = \frac{n!}{(n-n_{\text{max}})!} 
\end{split}
\label{eq: actspace}
\end{align}
A proper mapping between the integers in the action-space and the corresponding gate-qubit combination is achieved in two steps. The first value serves as an index for a list representing the gate set, thus selecting a specific gate. The second value indexes a listing of all qubit permutations possible, given distinct values for \(n\) and \(n_{\text{max}}\). Through this, it is decided which qubit combination the selected gate is applied on. In case the gate takes fewer qubits than present in the respective combination, the gate is simply applied on the first $n_{g}$ qubits of the permutation, while $n_{g}$ represents the number of qubits taken by the gate. Implementing the action space using the two-set architecture ensures that in a random sampling case, the selection of every gate and every combination is equally probable. On the contrary, utilizing just one set including all possible gate-qubit combinations in the first place, would lead to an unequal weighting of gate selection. This happens since gates taking a higher qubit number than others are applicable to a larger number of different qubit combinations and thus would appear more often in the set. Hence we chose the two-set architecture.

\subsection{Reward}
The quantum circuit environment comes with two different reward functions that are user selectable, a step-penalty reward (\cref{eq: reward2}) and a distance-based reward (\cref{eq: reward3}).

\begin{align}
    r_1 = \begin{cases}
         L - l -1   &\text{if} \quad  1 - F  <  SFE \\
        -1  &\text{otherwise}
        \end{cases}\label{eq: reward2}
\end{align}
Note that $l$ refers to the depth of the current quantum circuit, $L$ is the maximum circuit depth before terminating the episode, $F$ references the fidelity from \cref{eq: fidelity} and $SFE$ is the standard fidelity error describing the deviation of the fidelity $F$ from the value 1.

\begin{align}
    r_2 = \begin{cases}
         L - l -1   & \text{if} \quad 1 - F  <  SFE \\
        - \lfloor \frac{L}{2}  \rfloor \cdot (1-F)  & \text{if} \quad 1 - F  \geq  SFE  \;\land\;  l = 0 \\
        -1 & \text{otherwise}
        \end{cases}         \label{eq: reward3}
\end{align}
When comparing both equations, it becomes obvious that they are equivalent apart from the case when the episodes are finished without reaching the target. In the step-penalty reward equation, the agent just receives another $-1$ penalty, whereas in the distance reward equation, the final penalty is proportional to the distance of the current state to the target state $(1-F)$. Using this approach, the agent receives additional information about the closeness to the target, even if it was not able to reach it completely. This additional information may foster the learning procedure of the agent, especially when more complex targets are involved. 

\begin{figure}[htbp]
    \centering
    \includegraphics[width=\linewidth,keepaspectratio]{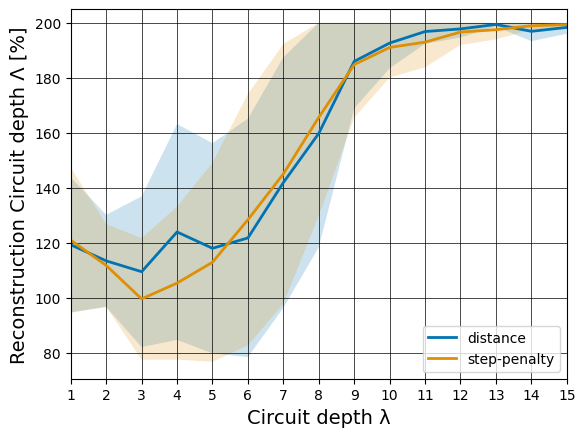}
    \caption{Comparison of two reward techniques: step-penalty and distance. Each data point represents the average performance of three runs, trained in a 2-qubit environment with varied target circuit-depth.} 
    \label{fig: rewardshaping}
\end{figure}
We evaluated both reward functions on 2-qubit-systems with targets characterized by circuit-depths $\lambda$ from 1 to 15, following the target-initialization algorithm (see \cref{subsec: envinit}). We executed 3 runs for every setting. Data points were obtained by averaging the reconstruction circuit-depth \(\Lambda\) (see \cref{subsec: metrik}) over the last 100 training episodes, with each agent undergoing identical training steps. Analysis of the reward technique comparison in \cref{fig: rewardshaping} reveals that while step-penalty and distance rewards exhibit similar behaviors, key differences emerge. Specifically, the step-penalty curve appears smoother and demonstrates a higher quality, correlating to a 5-20\% reduction in reconstructed circuit-depths \(\Lambda\) at lower \(\lambda \in \{2,3,4,5\}\). Owing to its robust performance and stability, especially at lower difficulty targets, the step-penalty technique was adopted as standard for all subsequent experiments.

\subsection{Target State Initialization}\label{subsec: envinit}
When initializing the environment, either a target quantum state or a circuit-depth $\lambda$ must be specified. If a target quantum state is set, a maximum depth \(L\) must also be defined, upon exceeding the current episode will terminate. We included this option to provide the possibility to apply agents on specific, fixed target states.

\begin{figure}[htb]
    \centering
    \includegraphics[width=\linewidth]{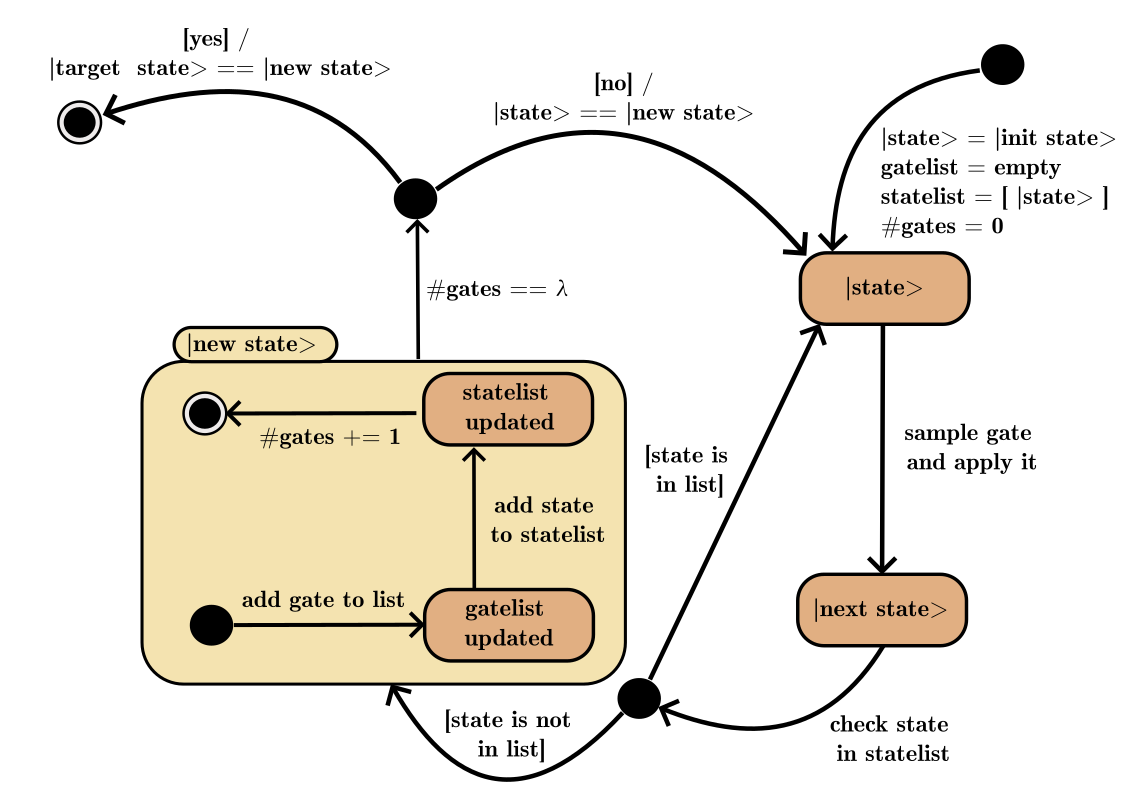}
    \caption{State diagram displaying the target state generation algorithm, while gate-list and state-list are implemented as actual lists and $\lambda$ is the circuit-depth parameter defining the absolute number of gates, which must be applied to get to the target. The algorithm starts at the upper right side of the figure.}
    \label{fig: targetgeneration}
\end{figure}
\bigskip

If no target parameter is provided, the circuit-depth parameter \(\lambda\) must be set, enabling the generation of a random target state per episode, as illustrated in \cref{fig: targetgeneration}. The parameter \(\lambda\) denotes the necessary quantum gate count, applied via a defined algorithm, to reach the target state. To ensure each gate meaningfully alters the circuit, a change condition is enforced for every additional gate, permitting its application only if the state change satisfies the condition $1-F \geq 0.001$, with $F$ corresponding to the fidelity. The change condition is then determined between the state after gate application and every state previously visited within the initialization procedure respectively. If one of the conditions is not satisfied then the gate is not applied. By this ineffective or neutralizing gate applications are avoided. For instance, with the Clifford+T gate set, redundancy can arise due to the self-inverse nature of Hadamard- and CNOT-gates. Utilizing this method mitigates cyclical patterns during target initialization and enhances the approximation accuracy of the actual minimal gate count to the provided \(\lambda\). However, exact equivalence is not assured. If a permitted gate cannot be applied after \(2 \times |G|\) consecutive tries, where \(|G|\) signifies the gate set cardinality, the generation process is reset to prevent stagnation at a specific state.

\subsection{Training Loop}
\begin{figure}[htbp]
    \centering
    \includegraphics[width=\linewidth]{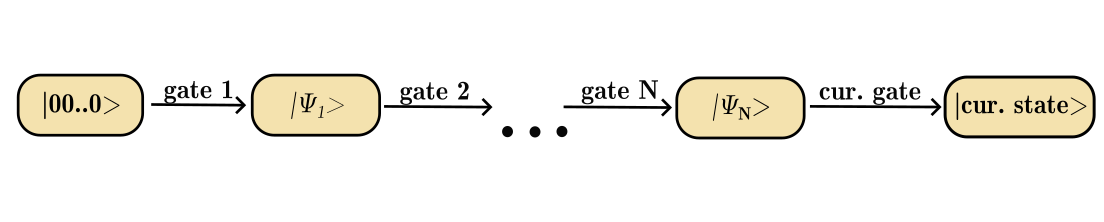}
    \caption{Schematic of the sequential application of the updated list of actions on the initial state $|00..0>$
    transforming it to the current state outputted by the environment.}
    \label{fig: gatesequence}
\end{figure}

Once the environment is initialized, the agent receives the initial quantum state and target state as first observation in the format specified in \cref{eq: obsspace}. The agent now chooses an action in the format specified in \cref{eq: actspace} based on the observation. This action corresponds to a gate applied to a specific combination of qubits. The environment appends the received action to the list of previously taken actions. The updated list of actions is then applied sequentially to form the respective quantum circuit present at the current step. Running the updated circuit then produces the succeeding observation equivalent to the next state of the environment. Following this procedure starting from a specified initial state (e.g. $|00...0\rangle$), the agent tries to apply a sequence of gates in order to get to the defined target state. This process is displayed in \cref{fig: gatesequence}. 
After the application of the chosen gate, the number of steps taken is increased by one, tracked by a step-counter variable $l$. In case $l$ reaches the maximal calculation length $L$, the episode is aborted. $L$, if not defined at the environments' initialization together with a specific target state, is determined by the otherwise given circuit-depth parameter $\lambda$ according to \cref{eq: maxcalclength}.
\begin{align}
     L = 2 \cdot \lambda \label{eq: maxcalclength}
\end{align} 
The other case in which the current episode is terminated occurs when the target state is reached. Hence one episode of the environment can be defined by taking steps starting from the initial state, either until the target state is reached or until the number of already taken steps equals the maximal calculation length $L$. When the episode is terminated in case no target state parameter is set, a new target is generated following the algorithm described in \cref{fig: targetgeneration} prior to the start of the next episode. Additionally, the current state is reset to the initial state and the step counter variable $l$ is set to 0 when a new episode is started.

\section{EXPERIMENTAL SETUP} \label{sec:experimental-setup}
In the following section we go into details about how we set up our experiments. We explain our choice of baseline algorithm and propose the reconstructed circuit depth as an evaluation metric specifically for the quantum circuit environment. Lastly, we elaborate on the training procedure and the used hyperparameters.

\subsection{Baselines}

In order to evaluate our environment designed for quantum circuit synthesis, we conducted tests using RL. Inspired by the approach of Gabor et al., we used Proximal Policy Optimization (PPO) and Advantage Actor-Critic (A2C) based agents as implemented in the stable-baselines framework, while adding a random agent for comparison reasons. Our aim was to identify and select the highest-performing algorithm among these, which was then utilized for our main experiments.

Our primary baseline, the random agent, serves as a rudimentary control, making arbitrary selections from the action space. This agent applies quantum gates onto the qubits without any informed guidance or strategy, providing a baseline performance metric which any proficient strategy should surpass.

Next, we implemented a PPO agent, which optimizes the policy by constraining the new policy to be close to the old policy~\cite{Schulman2017}. After performing a minor hyperparameter search, particularly focusing on the learning rate, the agent was evaluated on a 2-qubit circuit with varying circuit depths over three runs.

Likewise, A2C was evaluated, a synchronous, deterministic variant of A3C which uses advantage functions to reduce the variance of the policy gradient estimate~\cite{Mnih2016}. Following a similar experimental protocol as with PPO, it was subjected to a limited hyperparameter search, primarily adjusting the learning rate, and further evaluated under identical conditions on the 2-qubit circuit.


PPO distinctly outperformed A2C in synthesizing 2-qubit circuits across varied circuit depths, manifesting more consistent and proficient results over the three runs. The evaluations were quantitatively assessed based on the circuit-depth $\lambda$, ensuring a comprehensive appraisal across numerous scenarios and depths. Henceforth, due to its demonstrable superiority in our preliminary experiments, PPO was selected as the algorithm of choice for succeeding experiments and evaluations throughout our research.

\subsection{Reconstructed Circuit-Depth Metric}\label{subsec: metrik}
Establishing a performance metric is crucial to maintain a general comparability within the RL environment, particularly when involving diverse agents and parameterizations. Comparing different runs with varying configurations is difficult because the rewards scale with the used parametrization. To address this, we introduce the reconstructed circuit-depth metric, designed to normalize rewards with respect to the configuration, thereby facilitating a straightforward comparison across runs and giving a much more clean reading on the actual performance of the agent. By normalizing the number of gates used by the agent to recreate the target $n_{g}$ through circuit-depth $\lambda$ used for target initialization, a generally valid measure can be designed. This metric then represents how well the agent performed, normalized on the predefined circuit-depth. Defining the maximal calculation length $L$ according to \cref{eq: maxcalclength} and extracting the remaining calculation length of the current episode, $n_{g}$ can easily be derived using \cref{eq: takengatenumber}.
\begin{align}
    n_{g} =  L -l \label{eq: takengatenumber}
\end{align}
Following the intuitive setup of the general measure mentioned above, we define a metric called the reconstructed circuit-depth $\Lambda$ via \cref{eq: gatenorm}.
\begin{align}
     \Lambda[\%] = \frac {n_{g}}{\lambda} \cdot 100 \%\label{eq: gatenorm}
\end{align}
Achieving \(\Lambda = 100\%\) signifies that the agent has found a method to recreate the target with equal circuit-depth as the initial generation algorithm. The metric's limit extends to \(200\%\), considering that the maximum \(n_g\) value equals \(L\). Consequently, reconstructed circuit-depths ranging between \(100\%\) and \(200\%\) or below \(100\%\) indicate respectively longer or shorter gate sequences found by the agent compared to the intended gate sequence.

\section{Training and Hyperparameter Optimization}\label{sec: optimize}

To assure a robust and replicable training process, each model configuration was trained under consistent conditions on a Slurm cluster, utilizing Intel(R) Core(TM) i5-4570 CPU @ 3.20GHz and Nvidia 2060 and 1050 GPUs, for a substantial total of 1,700,000 steps (56667 - 1700000 episodes, dependent on the settings) per run. Three different seeds, namely $\{1, 2, 3\}$, were used for all experiments to obtain consistent and reliable results, outputting an average and a standard deviation for each data point. If not further specified, we use a $2$ qubit system with a circuit depth of $5$ and a SFE of $0.001$. Furthermore, we use an initial state of $|00...0>$, since it is the ground state of most quantum hardware and thus a prominent starting point~\cite{Schneider2022,Kaye2006,Blazina2005}.

A hyperparameter search was conducted, focusing on the learning rate, using a grid search approach across the following candidates: \(0.00001, 0.0001, 0.0003, 0.0005, 0.0007, 0.001, 0.01\). For the Proximal Policy Optimization (PPO) algorithm, a learning rate of \(0.001\) was identified as optimal and was subsequently utilized in all related experiments. Additionally, the clipping parameter for PPO was set to \(0.2\) to ensure stable and reliable policy updates. On the other hand, the Advantage Actor-Critic (A2C) algorithm demonstrated optimal performance with a learning rate of \(0.00001\).

This meticulous approach to training and hyperparameter optimization lays a solid foundation for the subsequent experiments, ensuring that the derived results and insights are both reliable and grounded in a systematic exploration of the model's parameter space.

\section{RESULTS} \label{sec:results}

A main objective of our project was the implementation of a RL environment capable of training RL agents on the DQCS problem. Another one was to study the DQCS itself. Therefore, we now conduct a variety of experiments to examine the task. For the setup used in the following experiments, see \cref{sec:experimental-setup}. 

A comparison of PPO and A2C used with the respective optimized settings showed a clear superiority of the PPO algorithm. We evaluated the last 100 episodes of the respective runs of the agents on 2-qubit targets with circuit-depth $\lambda = 5$. The reconstructed circuit-depth $\Lambda$ reached by the PPO agents was 112.5$\pm$35.3\%, while the A2C agents produced a $\Lambda$ of 195.3$\pm$6.5\%. In comparison the random baseline, corresponding to an untrained agent, exhibited a $\Lambda = 199.6\pm2.8$\%. Due to the better performance of PPO it was selected for all following experiments.

\subsection{Qubit -- Circuit-Depth Relationship} \label{sec: Difflandscape}
\begin{figure}[htbp]
    \centering
    \includegraphics[width=\linewidth,keepaspectratio]{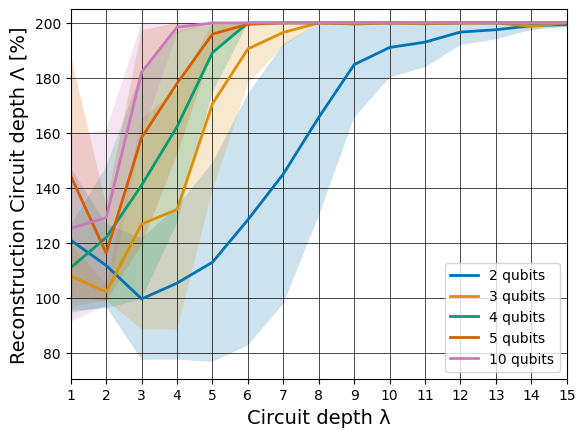}
    \caption{ Agents' performances on targets with different qubit numbers $n$ (2, 3, 4, 5, 10) while the circuit-depth $\lambda$ is varied (1-15). Every data point represents the average performance of 3 agents trained on systems possessing the respective $n$ and $\lambda$ settings.} 
    \label{fig: densegrid}
\end{figure}

In order to explore the complexity of the DQCS problem, we investigated the training of PPO-agents on various systems differing in their qubit numbers and circuit-depths $\lambda$. We trained agents on systems of 2, 3, 4, 5, and 10 qubits utilizing circuits of depths $\lambda$ from 1 to 15 respectively.
We obtained the data displayed in \cref{fig: densegrid} by averaging the reconstructed circuit-depth $\Lambda$ of the respective trained agents over the last 100 episodes of the training run.
Analyzing the curves, two main trends get apparent. The first observed trend is the increasing reconstructed circuit-depth ($\Lambda$) as the target circuit-depth ($\lambda$) rises, resembling a sigmoid curve. A higher target circuit-depth ($\lambda$) naturally demands more intricate solutions from the agent for their preparation. Starting from the left side of the plot, the reconstructed circuit-depth ($\Lambda$) ranges from 100\% to 140\% indicating scenarios where the agent's solutions resemble the paths used for target initialization. The curves exhibit a linear rise before eventually asymptotically converging towards a $\Lambda$ of 200\%. This indicates the agent's inability to recreate the target at this level of complexity.
The second trend involves a decrease in agent performance as the qubit number ($n$) increases, reflecting the increased complexity of the target state. This is evident from the steeper rise in the reconstructed circuit-depth ($\Lambda$) occurring at lower $\lambda$-values for targets with higher $n$ compared to those with lower $n$.
Of particular interest is the situation at $\lambda = 3$ and $n = 2$ (represented by the blue curve), where the curve tangentially intersects the 100\% mark. This scenario offers two possible explanations. First, it could imply that the agent perfectly recreates every target using an equivalent circuit-depth as applied during initialization. Alternatively, it suggests that the agent achieves the desired states with a lower circuit-depth and, consequently, fewer quantum gates than initially used to create the targets. Consequently, the average reconstruction circuit-depth ($\Lambda$) can reach 100\% or even drop below it. This observation, considering the advanced algorithm used for target initialization, provides evidence of the high level of optimization achieved by the trained agents.

\subsection{Benchmarking Analysis}
In the following section, we set up a benchmarking framework for the quality validation of the trained agents and for the comparison of different RL algorithms on the DQCS problem. We focus on the examination of comparatively simple systems for benchmarking, involving 2-qubit targets only. To facilitate the benchmarking, we conducted two different approaches. The first method measures the performance of an agent on a set of randomly generated targets of a certain circuit-depth $\lambda$, the other applies the agent on a set of specific well-known target states.

\subsubsection{Evaluation Levels}
The difficulty of the target circuit is dependent on two different factors, the number of qubits and the circuit-depth $\lambda$ used for target initialization. However, setting the qubit number to 2 leaves us with $\lambda$ as the only parameter. 
Based on the data obtained in \cref{sec: Difflandscape}, we are able to propose a splitting of the parameter-space of $\lambda$ [1,15] into three evaluation levels 'easy', 'medium', and 'hard'. According to \cref{fig: densegrid}, the first segment describes a development varying around a reconstructed circuit-depth of 110\%. The second region can be defined as an interval of linear rise and the last section is characterized by an asymptotic convergence against a reconstructed circuit-depth of 200\%. \cref{table: regimes} contains the definition of the derived levels.

\begin{figure}[htbp]
    \centering
    \includegraphics[width=\linewidth,keepaspectratio]{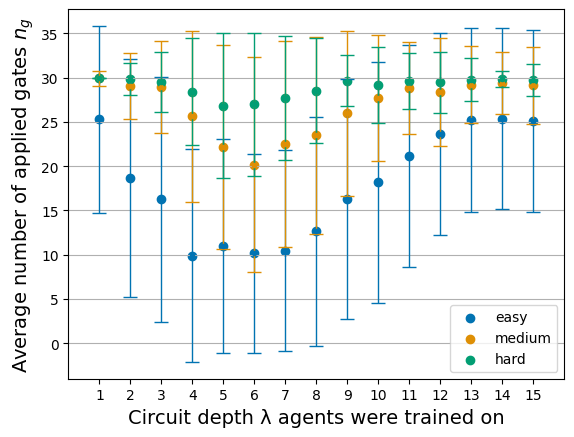}
    \caption{The average number of gates $n_{g}$ applied by PPO-agents trained on different circuit-depths $\lambda$, when applied on 100 random targets from the evaluation levels 'easy', 'medium' and 'hard' respectively.} 
    \label{fig: randombenchmark}
\end{figure}

\begin{table}[htbp]
\begin{center}
    \begin{tabularx}{\linewidth}{ |X|c| } 
         \hline
         Level & Set of included circuit-depths $\lambda$  \\ 
         \hline\hline
         easy &  \{1, 2, 3, 4, 5\} \\ 
         \hline
         medium &  \{6, 7, 8, 9, 10\} \\ 
         \hline
         hard &  \{11, 12, 13, 14, 15\} \\ 
         \hline
    \end{tabularx}
    \caption{The definition of the three evaluation levels 'easy', 'medium' and 'hard' within the circuit-depth $\lambda$ interval of $[1,15]$ }
    \label{table: regimes}
\end{center}
\end{table}
The data displayed in \cref{fig: randombenchmark} was obtained by applying agents trained on 2-qubit targets with different circuit-depths $\lambda$, ranging from 1 to 15 on 100 random targets of the respective evaluation level, while setting all other parameters to the standard values. For this experiment, the upper bound of possible applied actions, corresponding to the maximal calculation length $L$ was set to 30 steps. It's evident that each segment has been successfully solved by at least one agent with an average gate number below the maximum step count. When examining the bar chart, several trends become apparent. Firstly, there is a clear correlation between evaluation levels and the average gate number required for target preparation. Targets categorized as 'hard' typically demand more gates, on average, for their preparation compared to those categorized as 'medium' or 'easy.' This trend arises from the general need for more gates when dealing with higher circuit-depth ($\lambda$) targets. Another noteworthy observation pertains to the absence of a shift in peak performance towards agents trained on high circuit-depth ($\lambda$) targets when transitioning from 'easy' to 'medium' and 'hard' evaluation levels. Intuitively, one might expect such a shift, given that changing the evaluation level involves applying agents to target states with different average circuit-depths ($\lambda$). Consequently, it would be reasonable to anticipate a shift of the best-performing agents towards those trained on a circuit depth ($\lambda$) close to the average circuit-depth of the respective evaluation level. However, this expected correlation seems to be lacking, indicating a weak dependency between these variables. 

The experiment showed a high similarity of the best-performing agents for the respective evaluation levels. Specifically, agents trained with $\lambda$ values ranging from 4 to 7 exhibited notable performance. Agents with $\lambda$ values of 6 and 7 consistently ranked within the top 3 across all regions, while $\lambda = 5$ appeared in the top list of 2 sections. This shared performance trend may be attributed to the initial optimization, which emphasized agents with similar settings.\indent For the subsequent benchmarking of well-known states, we chose the best-performing agents on the respective evaluation levels (agents trained with $\lambda =$ 4, 5, and 6) as our candidates.

\subsubsection{Reconstructing Well-Known States}
\begin{figure}[htbp]
    \centering
    \includegraphics[width=\linewidth,keepaspectratio]{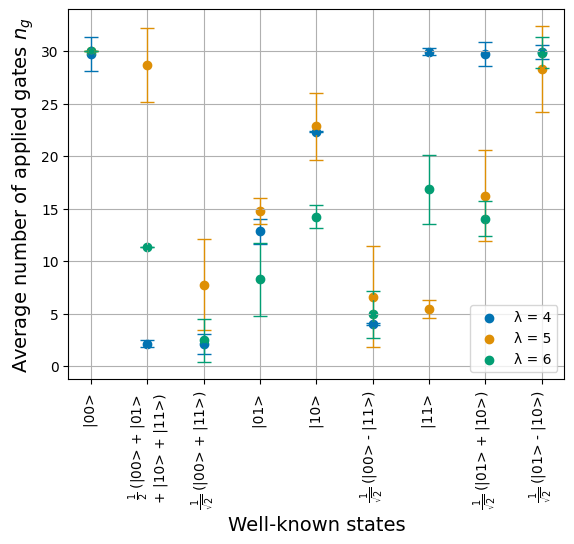}
    \caption{Performance of PPO-agents trained on circuit-depths $\lambda$ = 4,5 and 6 which are applied on the 9 states included in the well-known state set given in \cref{table : stateset}.} 
    \label{fig: testsetbenchmark}
\end{figure}

As mentioned earlier this benchmarking method again focuses on 2-qubit systems only. We composed the set out of states well-known in the quantum community, including the four basis states of the 2-qubit state-space $\ket{00},\ket{01},\ket{10}$ and $\ket{11}$, the completely mixed state $\frac{1}{2}(\ket{00} + \ket{01} + \ket{10} + \ket{11})$ and the four 2-qubit bell states $\frac{1}{\sqrt{2}}(\ket{00} + \ket{11})$, $\frac{1}{\sqrt{2}}(\ket{00} - \ket{11})$, $\frac{1}{\sqrt{2}}(\ket{01} + \ket{01})$ and $\frac{1}{\sqrt{2}}(\ket{01} - \ket{01})$. We divided the chosen benchmark states into subgroups representing different levels of evaluation, according to the minimal circuit-depth necessary to prepare the respective states. We obtained the minimal sequences starting from the ground state $\ket{00}$, using a brute-force searching algorithm trying all possible combinations of gates included in the Clifford+T gate set. According to the minimal circuit-depth, we divided the set into three subgroups (easy: 0-2 gates, medium: 3-4 gates, hard: $>5$ gates) The results of this classification are displayed in \cref{table : stateset}.\indent
\begin{table}[htb]
\begin{center}
    \renewcommand{\arraystretch}{1.5}
    \begin{tabularx}{\linewidth}{ |l|C|C| } 
         \hline
         State & Minimal circuit-depth  & Level \\
         \hline\hline
         $\ket{00}$ &  0 & easy \\
         \hline
         $\frac{1}{2}(\ket{00} + \ket{01} + \ket{10} + \ket{11})$  & 2 & easy \\
         \hline
         $\frac{1}{\sqrt{2}}(\ket{00} + \ket{11})$  & 2 & easy \\
         \hline 
         \hline 
         $\ket{01}$ & 4 & medium \\
         \hline
         $\ket{10}$ & 4 & medium \\
         \hline
         $\frac{1}{\sqrt{2}}(\ket{00} - \ket{11})$ & 4 & medium \\
         \hline
         \hline 
         $\ket{11}$ & 5 & hard \\
         \hline
         $\frac{1}{\sqrt{2}}(\ket{01} + \ket{10})$ & 5 & hard \\
         \hline
         $\frac{1}{\sqrt{2}}(\ket{01} - \ket{10})$ & 7 & hard \\
         \hline
    \end{tabularx}
    \caption{All states included in the 2-qubit set, divided into subgroups of different evaluation levels, listed together with the minimal number of quantum gates necessary for their preparation using only gates contained in the Clifford+T gate set. }
    \label{table : stateset}
\end{center}
\end{table}
Subsequently, we tested agents trained with $\lambda =$ 4, 5, and 6 on the designed set of states. The outcomes of this investigation are displayed in \cref{fig: testsetbenchmark}, while the targets are ordered according to the rise in their minimal required circuit-depth from left to right. An important fact getting evident from the displayed data is the presence of big performance variations between the agents when applied to certain targets. Further, the quality of the agents themselves varies strongly for different targets. This indicates a form of specialization of certain agents on specific targets.

For the state $\ket{00}$, which is equivalent to the initial state, all tested agents struggle to solve the task. It's essential to note that the environment doesn't check if the initial state matches the target initially. Hence, in cases where the target is already reached from the beginning, the agent must apply a gate that preserves the state, such as S-, T-, or an identity gate on any qubit. Despite these options comprising more than half of the available action-space, they are rarely selected, resulting in high $n_{g}$-values. This behavior is explained by the fact that agents are typically trained on targets different from the initial state, requiring gates that modify the current observation. Moreover, due to the target initialization algorithm's implementation, $\ket{00}$ is never a target during training when $\lambda \neq 0$. Consequently, agents are explicitly trained to avoid applying gates that would be useful in this scenario, leading to suboptimal performance on this specific task.

The $\lambda = 4$ agents being trained on a comparatively low circuit-depth $\lambda$ show the best results for relatively easy states like  $\frac{1}{2}(\ket{00} + \ket{01} + \ket{10} + \ket{11})$ and $\frac{1}{\sqrt{2}}(\ket{00} + \ket{11})$, while lacking in performance when applied to the targets of a higher minimal circuit-depth. On the other hand, agents trained with $\lambda$ = 5 and 6 are able to perform better on the targets of a higher minimal circuit-depth, showing strong performance in general. For the state $\frac{1}{\sqrt{2}}(\ket{01} - \ket{10})$ possessing the highest minimal circuit-depth, all agents show low potential. To analyze the generated solutions in more detail, the following \cref{table : circuitanalysis} contains the most promising quantum circuits created within this benchmark test. We obtained these circuits by identifying the $\lambda \; \in \; \{4,5,6\}$, which showed the lowest $n_{g}$ for a particular target and subsequently determined the top-performing agent with the respective $\lambda$. The most frequently generated circuit produced by this agent was then extracted for the respective target state. Since the states $\ket{00}$ and $\frac{1}{\sqrt{2}}(\ket{01} - \ket{10})$ lead to a wide distribution of created circuits without a clear favorite, they were excluded from this analysis. The 'Generation probability' describes the likelihood of occurrence of the specific circuit in percent. Considering the gates included in the displayed solutions, while bearing in mind the minimal circuit-depth (see \cref{table : stateset}), it becomes apparent that for all targets a minimal circuit was found. Further, designs like the circuit created for $\frac{1}{\sqrt{2}}(\ket{00} + \ket{11})$ are known to literature as well~\cite{Schneider2022}. The generation probabilities of the results show that the agents strictly specialized in the preparation of specific circuits, obtaining values of 96-100\% for all but one state. Conclusively it can be stated that the selected agents perform sufficiently on real-world examples as included in the set, creating highly optimized quantum circuits which indeed prepare the desired targets.

\begin{table*}[t]
\begin{center}
    \begin{tabularx}{\linewidth}{ |C|C|c| } 
         \hline
         State &  Generated circuit & Gen. probability [\%] \\ 
         \hline\hline
         \raisebox{-20pt}{$\frac{1}{2}(\ket{00} + \ket{01} + \ket{10} + \ket{11})$ }  & \raisebox{-20pt}{\begin{quantikz}[thin lines]&\gate{H}&&&&&&&&& \\ &\gate{H}&&&&&&&&&\end{quantikz}}& \raisebox{-20pt}{100}\\[40pt]
         \hline 
         \raisebox{-20pt}{$\frac{1}{\sqrt{2}}(\ket{00} + \ket{11})$}  & \raisebox{-20pt}{\begin{quantikz}[thin lines]&\gate{H}&\ctrl{1}&&&&&&& \\ &&\targ{}&&&&&&& \end{quantikz}} & \raisebox{-20pt}{100}\\[40pt]
         \hline 
         \raisebox{-20pt}{$\ket{01}$} & \raisebox{-20pt}{\begin{quantikz}[thin lines]&&&&&& \\ &\gate{H}&\gate{S}&\gate{S}&\gate{H}&& \end{quantikz}} & \raisebox{-20pt}{96}\\[40pt]
         \hline
         \raisebox{-20pt}{$\ket{10}$} & \raisebox{-20pt}{\begin{quantikz}[thin lines]&\gate{H}&\gate{S}&\gate{S}&\gate{H}&& \\&&&&&& \end{quantikz}} & \raisebox{-20pt}{98}\\ [40pt]
         \hline
         \raisebox{-20pt}{$\frac{1}{\sqrt{2}}(\ket{00} - \ket{11})$}  & \raisebox{-20pt}{\begin{quantikz}[thin lines]&\gate{H}&\gate{S}&\ctrl{1} &&&& \\ &&&\targ{}&\gate{S}&&& \end{quantikz}} & \raisebox{-20pt}{76}\\[40pt]
         \hline
         \raisebox{-20pt}{$\ket{11}$} & \raisebox{-20pt}{\begin{quantikz}[thin lines]&\gate{H}&\gate{S}&\gate{S}&\gate{H}&\ctrl{1}& \\&&&&&\targ{}& \end{quantikz} }& \raisebox{-20pt}{99}\\[40pt]
         \hline
         \raisebox{-20pt}{$\frac{1}{\sqrt{2}}(\ket{01} + \ket{10})$} & \raisebox{-20pt}{\begin{quantikz}[thin lines]&\gate{H}&\ctrl{1} &&\targ{}&& \\ &&\targ{}&\gate{H}&\ctrl{-1}&\gate{H}&\end{quantikz} }& \raisebox{-20pt}{96}\\[40pt]
         \hline
    \end{tabularx}
    \caption{States included in the 2-qubit benchmark set listed together with the most probable circuit designs created by the best-ranked agents.}
    \label{table : circuitanalysis}
\end{center}
\end{table*}

\section{CONCLUSION} \label{sec:conclusion}
In this work, we describe the successful implementation of a RL environment, enabling the training of RL agents on the DQCS problem. Relying on the utilization of gates from the Clifford+T gate set only, we facilitate the direct transfer of the synthesized quantum circuits onto real quantum devices.
While exploring the parameter-space of the DQCS problem, we demonstrated sufficient task-solving capabilities of the trained agents, regarding DQCS in a wide variety of settings. The tested parameter-space spanned different targets including systems exhibiting qubit counts of 2, 3, 4, 5, and 10, as well as circuits-depths ranging from 1 to 15 used for target initialization. Through the investigation of the agents' behavior on different DQCS systems, we discovered correlations between the target state parameters, qubit-count and circuit-depth, and the agents reconstructed circuit-depth. It got evident that the reconstructed circuit-depth increases, if the qubit count or the circuit-depth is raised. We found 2-qubit targets to be solvable by the trained agents for a variety of different circuit-depths. However, the preparation of targets characterized by a qubit number $>2$ still poses challenges. 

Future efforts could further optimize hyperparameters and refine the agent's training target-space to boost RL network performance. We also consider adopting a curriculum learning or a GAN approach, with a generative network as the learner and a discriminator for target setting. For quantum computers using non-Clifford+T gate sets, we'll integrate parameterized gates to devise advanced quantum circuits. This, however, demands the RL agent to adapt further. Beyond studying the DQCS issue, we've set benchmarks comparing RL algorithms. Our trained PPO agents performed on discrete tasks, revealing optimal circuit designs. We plan to expand these benchmarks for systems over 2 qubits using k-means clustering.

In summary, our findings demonstrate the applicability and potential of reinforcement learning in addressing the DQCS problem, highlighting the need for further research. Our approach represents a significant step towards fully automated quantum circuit synthesis, showcasing the effectiveness of RL methods in tackling this challenge.

\section*{ACKNOWLEDGEMENTS}
This work is part of the Munich Quantum Valley, which is supported by the Bavarian state government with funds from the Hightech Agenda Bayern Plus.

\bibliographystyle{apalike}
{\small
\bibliography{main}}

\end{document}